\documentstyle{article}

\makeatletter
\@addtoreset{equation}{section}
\makeatother
\def\*{\,\cdot\,}
\def\a{\alpha}
\def\b{\beta}
\def\c{\gamma}
\def\d{\delta}
\def\l{\lambda}

\def\ds{\displaystyle}

\def\vectr#1#2{\left(\begin{array}{ccc}&#1&\\&#2&\end{array}\right)}
\def\matr#1#2#3#4{\left(\begin{array}{rr}#1&#2\\#3&#4\end{array}
\right)}
\begin{document}

\begin{titlepage}

\title{Eigenvector and eigenvalue problem
for $3D$ bosonic model.}

\author{
I. G. Korepanov\\
South Ural State University, 76 Lenin av.,\\ Chelyabinsk 454080, \\
and Ural Division of International Institute \\ for Nonlinear
Research,
Ufa,Russia.\\
E-mail: igor@prima.tu-chel.ac.ru\\ \\
S. M. Sergeev\\
Branch Institute for Nuclear Physics, Protvino 142284, Russia.\\
E-mail: sergeev\_ms@mx.ihep.su
}

\date{February, 1998}

\maketitle

\abstract
{
In this paper we reformulate free field theory models defined on
the rectangular $D+1$ dimensional lattices as $D+1$ evolution
models. This evolution is in part a simple linear evolution on free
(``creation''  and ``annihilation'') operators.
Formal eigenvectors
of this linear evolution can be
directly constructed, and them play the role of
the ``physical'' creation and annihilation operators. These operators
being completed by a ``physical'' vacuum vector give the spectrum
of the evolution operator, as well as the trace of the evolution
operator give a correct expression for the partition function.
As an example, Bazhanov -- Baxter's
free bosonic model is considered.
}

\end{titlepage}
\newpage

\section{Introduction}

In this paper we investigate the simplest case of integrable
statistical mechanics and field theory models: the free models.

To fix some definitions and terminology, recall in few words, for
example,
scholar scalar free field theory in four dimensions.
The hamiltonian
\begin{equation}\label{4dham}
\ds
H_0(\pi,\phi)\;=\;
{1\over 2}\;\int\;d\vec x\;
( \pi(\vec x)^2 + (\vec\partial\phi(\vec x))^2 + m^2\phi(x)^2)\;,
\end{equation}
gives the evolution in physical time.
In the quantum field theory a matrix element of the evolution
operator can be presented in the form of the path integral
of a Gaussian exponent.
Hamiltonian (\ref{4dham}) can be diagonalized
by the Fourier transformation
\begin{equation}
\ds
\phi(\vec x)\;=\;
{1\over (2\pi)^{3/2}}\;\int\;
(a^+(-\vec k) + a(\vec k))\;\exp( i\vec k\vec x)\;
{d\vec k\over \sqrt{2\omega}}
\end{equation}
and
\begin{equation}
\ds
\pi(\vec x)\;=\;
{1\over (2\pi)^{3/2}}\;\int\;
(a^+(-\vec k) - a(\vec k))\;\exp( i\vec k\vec x)\;
i\;\sqrt{\omega\over 2}\;d\vec k\;,
\end{equation}
with the dispersion relation
\begin{equation}\label{4ddr}
\ds
\omega^2\;=\;\vec k^2\;+\;m^2\;.
\end{equation}
The hamiltonian becomes
\begin{equation}
\ds
H_0\;=\;\int\;\omega\;a^+(\vec k)\;a(\vec k)\;d\vec k\;,\;\;\;
[a(\vec k),a^+(\vec k')]\;=\;\delta(\vec k-\vec k')\;.
\end{equation}
The $n$ -- particle
eigenvectors of the hamiltonian and so of the evolution operator
are the Fock states
\begin{equation}
\ds
\prod_{j=1}^{n}\;a^+(\vec k_j)\;|0>\;,\;\;\;
H_0\;|0>\;=\;0\;.
\end{equation}
General free models on regular lattices have the same feature
as this scalar model: all them can be diagonalized
by proper Fourier transformations \cite{bs-3dffm}.
The differences between the scalar model and a general free theory
on the rectangular lattice are:
\begin{itemize}
\item on the lattice the main object is not a hamiltonian,
but an one-step evolution operator in the discrete time.
\item dispersion relations arising in the lattice models are
little bit more complicated then (\ref{4ddr}), and
they give nontrivial partition functions of the lattice
models endowed by a phase transition phenomena.
\end{itemize}

In the well known realm of the two dimensional models one could
mention
the free fermionic model (F.F.M.) as such case. The partition
function
for F.F.M. can be represented in a form of a path -- type integral
over
the Grassmanian variables. The partition function is a determinant
of large enough matrix which can be block -- diagonalized with a help
of the Fourier transformation, in this way the well known expression
for the partition function for finite lattice can be obtained.
\footnote{The same is valid for the $3D$ F.F.M. \cite{bs-3dffm}.}

An advantage of a $1+1$ evolution formulation for
the free bosonic and fermionic models is that one may avoid
the path -- type integration, considering the
evolution of the creation and annihilation operators.

It is very useful also to deal not with the usual transfer matrices
as
the evolution operators, but with the transfer matrices appearing in
the
diagonal sections of $2D$ lattice. Such matrices $U$ do not form
a commutative family, but they commute with the set of usual transfer
matrices $T$, this is provided by the Yang -- Baxter equation.

\begin{center}

\setlength{\unitlength}{0.25mm} 
page.
\thicklines
\begin{picture}(450,400)
\put(0,50){
\begin{picture}(450,300)
\put(0,100){\vector(1,0){350}}
\multiput(50,50)(50,0){6}{\line(0,1){100}}
\multiput(50,150)(100,0){3}{\vector(1,2){50}}
\multiput(100,150)(100,0){3}{\vector(-1,2){50}}
\put(60,110){$p$}\put(110,110){$q$}
\put(160,110){$p$}\put(210,110){$q$}
\put(260,110){$p$}\put(310,110){$q$}
\put(45,30){$\sigma_1$}\put(95,260){$\sigma_1'$}
\put(95,30){$\sigma_2$}\put(145,260){$\sigma_2'$}
\put(145,30){$\sigma_3$}\put(195,260){$\sigma_3'$}
\put(195,30){$\sigma_4$}\put(245,260){$\sigma_4'$}
\put(375,200){$\ds \rightarrow U({p\over q})$}
\put(375,100){$\ds \rightarrow T(p,q)$}
\end{picture}}
\put(120,0){Fig. 1. The configuration $U\*T$.}
\end{picture}

\vspace{0.5cm}

\end{center}

\begin{equation}
\ds
U({p\over q})\*T(p,q)\;=\;T(p,q)\* U({p\over q})\;.
\end{equation}
The graphic configuration is shown in Fig. 1.

Such formulation of a free model as $1+1$ evolution system with the
cross -- type evolution operators can be generalized easily to the
$(D+1)$ -- dimensional case with $(D+1)$ -- simplex equation
as the condition of the existence of a commuting $T$ -- type
operators family.
Moreover, for $D+1$ evolution system it is also easy to
construct eigenvectors for $U$ and so for $T$.
Note, the transfer matrices $U$ in $3D$
were investigated originally in Ref. \cite{korepanov-u}.
In this paper we deal with $2+1$ evolution models.

The paper is organized as follows. In section 2 an operator
formulation
of free models in terms of Gaussian operators is given.
In section 3 we define a $2+1$ evolution model on the
cagome lattice and describe a formal diagonalization of it.
Examples are given in section 4.
Most interesting example is connected with
Bazhanov -- Baxter's free bosonic model
\cite{bb-second}, and so section 4 is devoted to it
mainly.

\section{Formulation}

Let ${\cal A}$ be an algebra with a left ${\cal A}$-module $F$.
Let ${\cal A}$ as the linear space has a basis $e_\a$.
Let ${\cal G}({\cal A})$ be a group of linear automorphisms
of ${\cal A}$: for any $G\in {\cal G}({\cal A})$
\begin{equation}\label{g-on-a}
\ds
e'_\a\;=\;G\,\circ\,e_\a\; =  \; \sum_\b\; g_{\a,\b}\;e_\b\;,
\end{equation}
where $G$ is properly defined on $F$ so as
\begin{equation}\label{g-on-f}
\ds
G\,\circ\,e\;\stackrel{def}{=}\;G\* e \*G^{-1}\;.
\end{equation}
Call these $G$ the Gaussian operators on ${\cal A}$. To
exhibit the definition of $G$ via the matrix $g_{\a,\b}$ we will
use the universal notation $G=G(g)$. Also we will
omit the indices $\a$ denoting the basis of ${\cal A}$ via $\vec e$
so that
\begin{equation}
\ds
G(g)\,\circ\,\vec e\;=\;g\*\vec e\;.
\end{equation}

Consider now the tensor product of $\Delta$ copies of $F$:
\begin{equation}
\ds
F^{\otimes\Delta}\;=\;
\underbrace{F\otimes F\otimes ... \otimes F}_{\Delta\mbox{ times}}\;.
\end{equation}
This linear space is the left module of the direct sum of
$\Delta$ copies of ${\cal A}$.
Let
\begin{eqnarray}
\ds
\vec e_k\;=\;1\otimes 1\otimes \dots \otimes
& \!\!\!\!\!\vec e \!\!\!\!\!&
\otimes 1 \otimes \dots.
\nonumber\\
\ds & \!\!\!\!\! \downarrow \!\!\!\!\! &\nonumber\\
\ds & \!\!\!\!\! k \!\!\!\!\! &-th \mbox{ place}
\end{eqnarray}
The Gaussian operators from ${\cal G}(\sum\;{\cal A}_k)$,
\begin{equation}
\ds
G(g)\,\circ\,
\left(\begin{array}{ccc}
& \vec e_1 & \\ & \vec e_2 & \\ & \vdots & \\ &\vec e_\Delta &
\end{array}\right)\;=\;
\left(\begin{array}{cccc}
 g_{1,1} & g_{1,2} & \dots & g_{1,\Delta} \\
 g_{2,1} & g_{2,2} & \dots & g_{2,\Delta} \\
 \vdots  & \vdots  & \ddots & \vdots      \\
 g_{\Delta,1} & g_{\Delta,2} & \dots & g_{\Delta,\Delta}
\end{array}\right)\*
\left(\begin{array}{ccc}
& \vec e_1 & \\ & \vec e_2 & \\ & \vdots & \\ &\vec e_\Delta &
\end{array}\right),
\end{equation}
we suppose to be properly defined on $F^{\otimes\Delta}$.

Note that the matrices $g$ represent the antihomomorphism
of the group of the Gaussian operators $G(g)$:
\begin{equation}
\ds
G(g)\* G(g')\;=\; G(g'\* g).
\end{equation}

Suppose for some reason we consider the Gaussian operators of special
type:
operators which act nontrivially only on some part
of $F^{\otimes\Delta}$. Namely, let
\begin{equation}
\ds
h\;=\;\{k_1,k_2,...,k_M\}\;,
\end{equation}
then denote
\begin{equation}
\ds
G_h\;\equiv\;G_{k_1,k_2,...,k_M}
\end{equation}
an operator such that
\begin{equation}
\ds
G_h\,\circ\,\vec e_j\;=\;\vec e_j\;,\;\;\;
j\;\not\in h\;,
\end{equation}
while for $\vec e_k$ with $k\in h$
\begin{equation}\label{gh}
G_h\,\circ\,
\left(\begin{array}{ccc}
&\vec e_{k_1}& \\
&\vdots&\\
&\vec e_{k_M}&\end{array}\right)\;=\;
\left(\begin{array}{ccc}
g_{k_1,k_1} & \dots & g_{k_1,k_M} \\
\vdots & \ddots & \vdots \\
g_{k_M,k_1} & \dots & g_{k_M,k_M}\end{array}\right)\*
\left(\begin{array}{ccc}
&\vec e_{k_1}& \\
&\vdots&\\
&\vec e_{k_M}&\end{array}\right).
\end{equation}

Such operators appear, for example, when one considers the
$D$-simplex
equation:
\begin{equation}\label{dsimpl}
\ds
G_{h}\*G'_{h'}\*\dots\*G^{(D)}_{h^{(D)}}\;=\;
G^{(D)}_{h^{(D)}}\*\dots\*G'_{h'}\*G_{h}\;,
\end{equation}
where each $h^{(m)}$ consists on $D$ indices, the whole number of the
indices in (\ref{dsimpl}) is $\ds {D(D+1)\over 2}$, and denoting
\begin{equation}
\ds
h\;=\;\{k_1,...,k_D\}\;\;\Leftrightarrow\;\; k_n\;=\;(h)_n\;,
\end{equation}
then
\begin{equation}
\ds
(h^{(m)})_n\;=\;(h^{(n)})_{m+1}\;.
\end{equation}
Note, $h\;=\;h^{(0)}$.

All operators $G_h$ are
defined in $\mbox{End }\ds F^{\otimes\Delta}$, (\ref{dsimpl}) is
the usual $D$-simplex equation for the operators acting in the tensor
product. But (\ref{dsimpl}) implies also the corresponding
equation for $g_h$:
\begin{equation}\label{dsimplsum}
\ds
g^{(D)}_{h^{(D)}}\*\dots\*g'_{h'}\*g_{h}\;=\;
g_{h}\*g'_{h'}\*\dots\*g^{(D)}_{h^{(D)}}\;,
\end{equation}
and the main difference between (\ref{dsimpl}) and (\ref{dsimplsum})
is that $g_h$-s act on the direct sum of linear spaces:
$\ds g_h\;\in\;\mbox{ End }\ds \sum_k\; {\cal A}_k$.
Such kind of $D$-simplex-type equations was investigated in Refs.
\cite{korepanov-diss}.
$\{\vec e_k\}$ is the formal basis, and $G(g_h)$-s are defined
on the vectors
\begin{equation}
\ds
\omega\;=\;\sum_{k=1}^{\Delta}\; \vec x_k^T\*\vec e_k \;.
\end{equation}
Thus there defined the co-action of $G(g)$ on the amplitudes $\vec
x_k$:
\begin{equation}
\ds
\vec x_k\;\;\mapsto\;\; \vec x_k^\prime\;=\; (g^T\* \vec x)_k
\end{equation}
due to
\begin{equation}
\ds
G(g)\,\circ\,\omega\;=\;\sum_k\; (g^T\*\vec x)_k^T\*\vec e_k\;.
\end{equation}

These simple facts allows one to formulate and solve eigenvalue and
eigenvector problem of a special type. This will be done in the next
section.

\section{$2+1$ evolution}

In this section we'll formulate some $2+1$ evolution problem.
The formulation
is rather dimension independent, as well as the solution.
We choose the dimension $2+1$ because this is not so convenient as
$1+1$ evolution problem and is not so cumbersome as an arbitrary
$D+1$ case. Also the examples we'll give belong to the $2+1$ case.

Suppose we are intending to investigate a model (statistical
mechanics
or field theory) defined on the cubic lattice: the cites of the
lattice are
\begin{equation}
\ds
\vec n\;=\; n_x\;\vec e_x\;+\;n_y\;\vec e_y\;+\;n_z\;\vec e_z,
\end{equation}
where $\vec e_x,\vec e_y,\vec e_z$ are orthogonal normal basis in
$E_3$,
$n_x,n_y,n_z\in {\cal Z}_N$, $N$ is the size of the lattice, the
periodical boundary conditions assumed. Instead of
dealing with the usual layer -- to -- layer transfer matrices, i. e.
section of the lattice between any
$n_z=k$ and $n_z=k+1$, consider
the section of the $3D$ lattice between $n_x+n_y+n_z=m$ and
$n_z+n_y+n_z=m+1$. Call this section $n_x+n_y+n_z=m$ as $Z_m$
(do not mix this with the ring of residues) and define
the evolution as the transition $...\; Z_m\mapsto Z_{m+1}\;...\;$.
Such evolution problem was suggested in Refs.
\cite{korepanov-u}.

Consider now the section, e. g. $Z_0$, in details.
The two - dimensional oriented cagome -- type lattice $Z_0$ is
shown in Fig. 2.
$Z_0$ consists of the oriented triples of the vertices,
\begin{equation}
\ds
Z_0\;=\;\{\;\mbox{Triples}\;\;
[\;V_{\a_i,\b_j}\;,\;V_{\a_i,\c_k}\;,\;V_{\b_j,\c_k}\;]\;\}\;,
\end{equation}
where the numbers are chosen so as $i+j+k=0$, and $V_{\a,\b}$ stands
for a vertex at the intersection of the lines $\a$ and $\b$ etc.

\begin{center}

\setlength{\unitlength}{0.25mm} 
page.
\thicklines
\begin{picture}(450,550)
\put(0,50){
\begin{picture}(450,450)
\multiput(0,150)(0,50){3}{\vector(1,0){450}}
\multiput(100,0)(-50,50){3}{\vector(1,1){350}}
\multiput(350,0)(50,50){3}{\vector(-1,1){350}}
\put(460,245){$\c_{k-1}$}\put(460,195){$\c_k$}\put(460,145){$\c_{k+1}
$}
\put(355,460){$\b_{j+1}$}\put(405,410){$\b_j$}\put(455,360){$\b_{j-1}
$}
\put(-10,360){$\a_{i-1}$}\put( 40,410){$\a_i$}\put(
90,460){$\a_{i+1}$}
\end{picture}}
\put(120,0){Fig. 2. A fragment of the lattice $Z_0$.}
\end{picture}

\vspace{0.5cm}

\end{center}

Fix some notations for the sake of simplicity and shortness.
First, introduce the $3D$ translations as the re-enumeration
operators:
\begin{eqnarray}
\ds
\tau_z \,\circ\, [V_{\a_i,\b_j},V_{\a_i,\c_k},V_{\b_j,\c_k}] & = &
[V_{\a_{i+1},b_{j}},V_{\a_{i+1},\c_{k}},V_{\b_{j},\c_{k}}]\;,
\nonumber\\
\ds
\tau_y \,\circ\, [V_{\a_i,\b_j},V_{\a_i,\c_k},V_{\b_j,\c_k}] & = &
[V_{\a_{i},b_{j+1}},V_{\a_{i},\c_{k}},V_{\b_{j+1},\c_{k}}]\;,
\nonumber\\
\ds
\tau_x \,\circ\, [V_{\a_i,\b_j},V_{\a_i,\c_k},V_{\b_j,\c_k}] & = &
[V_{\a_{i},b_{j}},V_{\a_{i},\c_{k+1}},V_{\b_{j},\c_{k+1}}]\;.
\end{eqnarray}
Pure $2D$ translations on the kagome lattice are
\begin{equation}
\ds
p_\a\;=\;\tau_y\;\tau_x^{-1}\;,\;\;\;
p_\b\;=\;\tau_z\;\tau_x^{-1}\;,\;\;\;
p_\c\;=\;\tau_z\;\tau_y^{-1}\;,
\end{equation}
so as
\begin{equation}
\ds p_\b\;=\;p_\a\* p_\c\;.
\end{equation}
Thus the position of any triple $i,j,k\;:\;i+j+k=0$ can be noted by
\begin{equation}\label{point}
\ds
p\;=\;p_\a^{-k}\* p_\c^{i}
\end{equation}
because of
\begin{equation}
\ds
p\,\circ\,
[\,V_{\a_0,\b_0}\,,\,V_{\a_0,\c_0}\,,\,V_{\b_0,\c_0}\,]
\;=\;
[\,V_{\a_i,\b_j}\,,\,V_{\a_i,\c_k}\,,\,V_{\b_j,\c_k}\,]\;.
\end{equation}
Note, by this notation we chose $\a_0,\b_0,\c_0$ to be special point
in the frame of reference.

Let the sizes of the lattice be $N$ in each direction:
\begin{equation}
\ds
p_\a^N\;=\;p_\b^N\;=\;p_\c^N\;=\;1\;,
\end{equation}
so that the total number of the vertices is $\Delta \;=\; 3\,N^2$.

\bigskip

Assign now to each vertex $V$ the algebra ${\cal A}_V$ with
the basis $\vec e_V$ as well as its module $F_V$.
Then for a triple $[V_{\a,\b},V_{\a,\c},V_{\b,\c}]$ introduce
a (Gaussian) operator $R$:
\begin{equation}\label{rmap}
\ds
R\;:\; F_{V_{\a,\b}}\otimes F_{V_{\a,\c}}\otimes F_{V_{\b,\c}}\;\;
\mapsto\;\;
F_{V_{\a,\b}}\otimes F_{V_{\a,\c}}\otimes F_{V_{\b,\c}}\;,
\end{equation}
where it is supposed that in the right hand side the orientation
of the vertices $[V_{\a,\b},V_{\a,\c},V_{\b,\c}]$ is changed as it is
shown in Fig. 3.

\begin{center}

\setlength{\unitlength}{0.25mm} 
page.
\thicklines
\begin{picture}(450,300)
\put(25,50){
\begin{picture}(400,200)
\put(25,50){\vector(1,0){150}}
\put(40,30){\vector(1,2){75}}
\put(160,30){\vector(-1,2){75}}
\put(50,50){\circle*{10}}
\put(100,150){\circle*{10}}
\put(150,50){\circle*{10}}
\put(75,180){$\a$}\put(120,180){$\b$}
\put(180,60){$\c$}
\put(225,150){\vector(1,0){150}}
\put(290,30){\vector(1,2){75}}
\put(310,30){\vector(-1,2){75}}
\put(250,150){\circle*{10}}
\put(300,50){\circle*{10}}
\put(350,150){\circle*{10}}
\put(220,180){$\a$}\put(375,180){$\b$}
\put(380,130){$\c$}
\put(190,90){$\ds\mapsto$}
\put(190,115){$\ds R$}
\end{picture}}
\put(150,0){Fig. 3. The action of $R$.}
\end{picture}

\vspace{0.5cm}

\end{center}

To distinguish the configurations call the left hand side one
as L.H.S. triple and the right hand side -- as R.H.S. triple.
Consider now the evolution of $Z_0$ generated by applying $R$-s to
all L.H.S. triples of $Z_0$. As the result we obtain $Z_1$
formed by other L.H.S. triples:
\begin{equation}
\ds
Z_1\;=\;\{\;\mbox{L.H.S. Triples}\;\;
[\;V_{\a_i,\b_j}\;,\;V_{\a_i,\c_k}\;,\;V_{\b_j,\c_k}\;:\;
i+j+k=1]\;\}\;.
\end{equation}
Applying $R$-s to new L.H.S. triples repeatedly, one obtains
\begin{equation}
Z_0\rightarrow
Z_1\rightarrow
Z_2\rightarrow
Z_3\rightarrow ...
\end{equation}
where
\begin{equation}
\ds
Z_m\;=\;\{\mbox{ L.H.S. Triples }
[\;V_{\a_i,\b_j}\;,\;V_{\a_i,\c_k}\;,\;V_{\b_j,\c_k}\;:\;i+j+k=m\;]\;
\}\;.
\end{equation}
For $U_m\;:\;Z_{m}\;\mapsto\;Z_{m+1}$ one may write the following
expression :
\begin{equation}
\ds
U_m\;=\;\prod_{i+j+k=m}\;
R_{\ds V_{\a_i,\b_j}, V_{\a_i,\c_k}, V_{\b_j,\c_k}}\;.
\end{equation}

\bigskip

Introduce shorter notations for $\vec e_V$ and $F_V$ assigned to
the vertices $V_{\a_i,\b_j}$, $V_{\a_i,\c_k}$ and $V_{\b_j,\c_k}$.
Namely, mark everything for $V_{\a,\b}$ by $x$, for $V_{\a,\c}$ -- by
$y$,
and $V_{\b,\c}$ -- by $z$, and the position of a triple
$[V_{\a,\b},V_{\a,\c},V_{\b,\c}]$ -- by its $p$ as in Eq.
(\ref{point}):
\begin{eqnarray}
&\ds
V_{\a_i,\b_j}\;=\;p_\a^j\* p_\b^i\,\circ\, V_{\a_0,\b_0}\;=\;
V_x(p_\a^j p_\b^i)\;\leftrightarrow\;&\nonumber\\
&\ds
\vec e_x(p_\a^j p_\b^i)\;,\;\;\;
F_x(p_\a^j p_\b^i)\;,
&\\
&\ds
V_{\a_i,\c_k}\;=\;p_\a^{-k}\* p_\c^i\,\circ\, V_{\a_0,\c_0}\;=\;
V_y(p_\a^{-k} p_\c^i)\;\leftrightarrow\;&\nonumber\\
&\ds
\vec e_y(p_\a^{-k} p_\c^i)\;,\;\;\;
F_y(p_\a^{-k} p_\c^i)\;,
&\\
&\ds
V_{\b_j,\c_k}\;=\;p_\b^{-k}\* p_\c^{-j}\,\circ\, V_{\b_0,\c_0}\;=\;
V_z(p_\b^{-k} p_\c^{-j})\;\leftrightarrow\;&\nonumber\\
&\ds
\vec e_z(p_\b^{-k} p_\c^{-j})\;,\;\;\;
F_z(p_\b^{-k} p_\c^{-j})\;,
&
\end{eqnarray}
With these notations the action of $R$, Eq. (\ref{rmap}), looks
like
\begin{equation}
\ds
R\;:\;F_{x}(p)\otimes F_{y}(p)\otimes F_{z}(p)\;\mapsto\;
F_{x}(\tau_x p)\otimes F_{y}(\tau_y p)\otimes F_{z}(\tau_z p)\;.
\end{equation}
This situation is shown in Fig. 4, analogous to Fig. 3.

\begin{center}

\setlength{\unitlength}{0.25mm} 
page.
\thicklines
\begin{picture}(450,300)
\put(00,50){
\begin{picture}(200,200)
\put(  10 ,  70 ){\vector(1,0){180}}
\put(  40 ,  10 ){\vector(1,2){90}}
\put( 160 ,  10 ){\vector(-1,2){90}}
\put(  70 ,  70 ){\circle*{10}}
\put( 130 ,  70 ){\circle*{10}}
\put( 100 , 130 ){\circle*{10}}
\put(  95 ,  85 ){$p$}
\put( 110 , 125 ){$x$}
\put( 135 ,  80 ){$y$}
\put(  60 ,  80 ){$z$}
\end{picture}}
\put(250,50){
\begin{picture}(200,200)
\put(  10 , 130 ){\vector(1,0){180}}
\put(  70 ,  10 ){\vector(1,2){90}}
\put( 130 ,  10 ){\vector(-1,2){90}}
\put(  70 , 130 ){\circle*{10}}
\put( 130 , 130 ){\circle*{10}}
\put( 100 ,  70 ){\circle*{10}}
\put( 110 ,  65 ){$x$}\put(75,140){$y$}\put(120,140){$z$}
\put(  85 ,  10 ){$\tau_x\,p$}
\put( 160 , 150 ){$\tau_y\,p$}
\put(  10 , 150 ){$\tau_z\,p$}
\end{picture}}
\put(210,140){\begin{picture}(30,50)
\put(10,10){$\mapsto$}
\put(15,30){$U$}
\end{picture}}
\put(150,0){Fig. 4. The action of $R$.}
\end{picture}

\vspace{0.5cm}

\end{center}

Let now all operators $R$ coincide:
\begin{equation}
\ds
R_{V_x(p),V_y(p'),V_z(p'')}\;=\;
R(\vec e_x(p),\vec e_y(p'),\vec e_z(p''))
\end{equation}
with the same operator function $R$. Then all $U_m$ differ only by
some
re-enumeration of the vertices. Choose the re-enumeration operator
$\tau\;=\;\tau_x$:
\begin{eqnarray}
\ds \tau\,\circ\, V_x(p) & = & \ds V_x(p)\;,\nonumber\\
\ds \tau\,\circ\, V_y(p) & = & \ds V_y(p_\a^{-1}p)\;,\nonumber\\
\ds \tau\,\circ\, V_z(p) & = & \ds V_z(p_\b^{-1}p)\;.
\end{eqnarray}
Then, obviously,
\begin{equation}
U_m\;=\;\tau^m\,\circ\,U_0\;=\;
\tau^m\* U_0\*\tau^{-m}\;.
\end{equation}
The $m$ -- layers evolution operator
$U^{(m)}\;:\;Z_0\;\mapsto\; Z_m$,
\begin{equation}
\ds
U^{(m)}\;=\;
U_{m-1}\*U_{m-2}\*...\*U_1\*U_0\;=\; \tau^m\*(\tau^{-1}\* U_0)^m\;.
\end{equation}
Thus the universal one step evolution operator
\begin{equation}
\ds
U\;=\;\tau^{-1}\* U_0\;\;:\;\;Z_0\;\mapsto\;Z_0
\end{equation}
arises.

\bigskip
\noindent
Now we are going to diagonalize $U$.

\bigskip
\noindent
Let further $R$-s (and so $U$) are indeed Gaussian operators.
Investigate
the action of $R$ and $U$ on the linear space $\sum\;{\cal A}_k$ of
\begin{equation}
\ds
\vec e_x(p)\;\,\;\; \vec e_y(p)\;,\;\; \vec e_z(p)\;\;.
\end{equation}
Let the nontrivial part of
\begin{equation}
\ds
r\in {\cal G}({\cal A}_x\oplus {\cal A}_y\oplus {\cal A}_z)\;,
\end{equation}
$R=R(r)$, is given by $3\times 3$ block -- matrix
\begin{equation}
\ds
R(r)\,\circ\,
\left(\begin{array}{ccc}
& \vec e_x &\\
& \vec e_y &\\
& \vec e_z &\end{array}\right)
\;=\;
\left(\begin{array}{ccc}
r_{x,x}, & r_{x,y}, & r_{x,z} \\
r_{y,x}, & r_{y,y}, & r_{y,z} \\
r_{z,x}, & r_{z,y}, & r_{z,z} \end{array}\right)\*
\left(\begin{array}{ccc}
& \vec e_x &\\
& \vec e_y &\\
& \vec e_z &\end{array}\right).
\end{equation}
The space $\sum_p\; {\cal A}_x(p)\oplus {\cal A}_y(p)\oplus {\cal
A}_z(p)$
consists of the vectors
\begin{equation}
\ds
\omega\;=\;
\sum_P\; \vec x_p^T\*\vec e_x(p)\;+\;\vec y_p^T\*\vec e_y(p)\;+\;
\vec z_p^T\*\vec e_z(p)\;,
\end{equation}
where the amplitudes $\vec x_p,\vec y_p,\vec z_p$ may be chosen so
that
\begin{equation}
\ds
U\,\circ\,\omega\;=\;\l\;\omega\;.
\end{equation}

Obviously,
\begin{equation}
p_\a\* U\;=\; U\* p_\a\;,\;\;\;\;
p_\b\* U\;=\; U\* p_\b\;.
\end{equation}
Hence
\begin{eqnarray}
&\ds
p_\a\,\circ\,\omega\;\equiv\;\sum_p\;\vec x_p^T\*\vec
e_x(p_\a\,p)\;+\;...
&\nonumber\\
&\ds =\;\sum_p\;\vec x_{p_\a^{-1}p}^T\*\vec e_x(p)\;+\;...
\;=\;k_\a\;\omega\;,&
\end{eqnarray}
and
\begin{equation}
\ds p_\b\,\circ\,\omega\;=\;k_\b\;\omega\;,
\end{equation}
$k_\a^N\;=\;k_\b^N\;=\;1$. Thus
\begin{equation}
\ds
[\vec x,\vec y,\vec z]_{p_\a^{-1} p}\;=\;
k_\a\;[\vec x,\vec y,\vec z]_{p}\;,\;\;\;\;
[\vec x,\vec y,\vec z]_{p_\b^{-1} p}\;=\;
k_\b\;[\vec x,\vec y,\vec z]_{p}\;.
\end{equation}
Consider now the co-action of $U$ on the amplitudes (see Fig. 5):
with the eigenvalue $\lambda$
\begin{eqnarray}
\ds
\vec x^T_p\;\lambda & = &
\vec x^T_p\* r_{x,x}\;+\;\vec y^T_p\* r_{y,x}\;+\;\vec z^T_p\*
r_{z,x}\;,
\nonumber\\
\ds
\vec y^T_p\;\lambda & = &
\vec x^T_{p_\a^{-1}p}\* r_{x,y}\;+\;
\vec y^T_{p_\a^{-1}p}\* r_{y,y}\;+\;
\vec z^T_{p_\a^{-1}p}\* r_{z,y}\;,
\nonumber\\
\ds
\vec z^T_p\;\lambda & = &
\vec x^T_{p_\b^{-1}p}\* r_{x,z}\;+\;
\vec y^T_{p_\b^{-1}p}\* r_{y,z}\;+\;
\vec z^T_{p_\b^{-1}p}\* r_{z,z}\;.
\end{eqnarray}

\begin{center}

\setlength{\unitlength}{0.25mm} 
page.
\thicklines
\begin{picture}(450,250)
\put(0,50){
\begin{picture}(200,200)
\put(50,50){\line(1,0){100}}
\put(50,50){\line(1,2){50}}
\put(150,50){\line(-1,2){50}}
\put(50,50){\circle*{10}}
\put(150,50){\circle*{10}}
\put(100,150){\circle*{10}}
\put(150,20){$\ds\vec y_p$}
\put(90,170){$\ds\vec x_p$}
\put(30,20){$\ds\vec z_p$}
\end{picture}}
\put(250,50){
\begin{picture}(200,200)
\put(50,50){\line(1,0){100}}
\put(50,50){\line(1,2){50}}
\put(150,50){\line(-1,2){50}}
\put(50,50){\circle*{10}}
\put(150,50){\circle*{10}}
\put(100,150){\circle*{10}}
\put(150,20){$\ds\vec y'_{p_\a^{-1}p}$}
\put(90,170){$\ds\vec x'_p$}
\put(30,20){$\ds\vec z'_{p_\b^{-1}p}$}
\end{picture}}
\put(220,150){$\mapsto$}
\put(220,165){$U$}
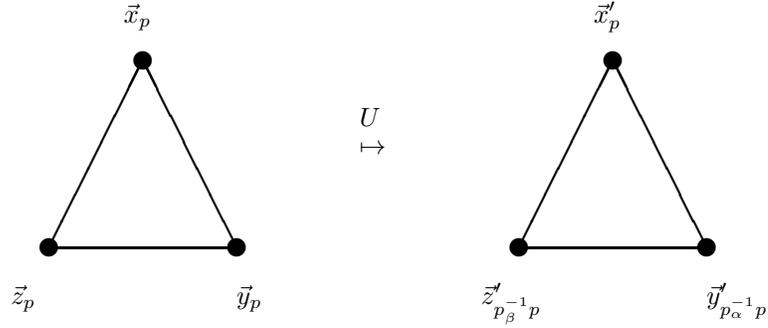
\put(120,0){Fig. 5. Transformation of the amplitudes. }
\end{picture}

\vspace{0.5cm}

\end{center}

Take into account the translation symmetry, then the finite
eigenvector and eigenvalue problem arises:
\begin{equation}
\ds
\left(\begin{array}{ccc}
r_{x,x}-\l, & r_{x,y}, & r_{x,z} \\
r_{y,x}, & r_{y,y}-\l k_\a^{-1}, & r_{y,z} \\
r_{z,x}, & r_{z,y}, & r_{z,z}-\l k_\b^{-1}
\end{array}\right)^T\*
\left(\begin{array}{ccc}
& \vec x_p & \\ & \vec y_p & \\ & \vec z_p &
\end{array}\right)\;=\;0\;.\label{eigen}
\end{equation}
Solving this problem, one obtains the eigenvalues of $U$ as the roots
of the characteristic polynom (the dispersion relation)
\begin{equation}
\ds
\chi(\l, \l k_\a^{-1},\l k_\b^{-1} |
r)\;=\;0\;\;\;\Leftrightarrow\;\;\;
\l_i\;=\;\l_i(k_\a,k_\b)\;,\;\;\; i=1,...,3\mbox{dim}{\cal A}\;,
\end{equation}
where
\begin{equation}\label{chp}
\ds
\chi(\l_1,\l_2,\l_3|r)\;=\;\mbox{det}\left(
\left(\begin{array}{ccc} \l_1 & 0 & 0 \\ 0 & \l_2 & 0 \\ 0 & 0 & \l_3
\end{array}\right) \;-\; r\right)\;.
\end{equation}
Let further
\begin{equation}
|\Omega>\;\in\; F^{\otimes \Delta}
\end{equation}
be the vacuum vector for $U$, i. e.
\begin{equation}
\ds
U\*|\Omega>\;=\;|\Omega>
\end{equation}
and $|\Omega>$ is the cyclic vector in $F^{\otimes\Delta}$.
Suppose $\vec e$ are free elements of ${\cal A}$, then the set of
eigenvectors of $U$ is the following set of the Fock -- type vectors
\begin{equation}\label{fock}
\ds
|n_1,...,n_\Delta>\;=\;
\omega_{\l_1}^{n_1}\*...\*\omega_{\l_\Delta}^{n_\Delta}\*|\Omega>
\end{equation}
with the eigenvalues
\begin{equation}
\ds
U\*|n_1,...,n_\Delta>\;=\;
\l_1^{n_1}\;...\;\l_\Delta^{n_\Delta}\;|n_1,...,n_\Delta>\;.
\end{equation}
Then formally
\begin{equation}
\ds
\mbox{ Trace } U^M\;=\;
\prod_\l\; {1\over 1-\l^M}\;,
\end{equation}
so for the partition function per cite $z_0(r)$
the bulk free energy $k_0(r)=\log z_0(r)$ in the
thermodynamic limit is given by
\begin{eqnarray}
\ds
k_0(r) & = & \ds-\mbox{ Lim }_{N,M\rightarrow\infty}\;
{1\over N^2\,M}\;
\sum_{k_a,k_b=0}^{N-1}\;\log\prod_{j}\;( 1-\lambda_j^M(k_\a,k_\b))
\nonumber\\
& = & \ds
-{1\over (2\pi)^3}\;\int\int\int_0^{2\pi}\;
d\phi_1\;d\phi_2\;d\phi_3\;
\log\; \chi(e^{i\phi_1},e^{i\phi_2},e^{i\phi_3}|r)\;.\label{integral}
\end{eqnarray}
In general vectors (\ref{fock}) are not all independent,
in this case one should take into account the difference
between ``the number of the degrees of freedom'' and the
dimension of ``phase space'' multiplying (\ref{integral})
by an appropriate constant $k_0 \mapsto k=\eta\,k_0$.
More precisely, for a given
representation of $\vec e_k$ on $F$ one should
extract from the set of the eigenvectors
$\omega_\lambda=\omega_i(k_\alpha,k_\beta)$ the operators,
cyclic with respect to $|\Omega>$, and sum the power series
of their eigenvalues.

\vspace{0.5cm}
\noindent
Thus for a Gaussian local evolution operators
one can easily construct the eigenvectors as well as the free energy
for the thermodynamic limit.
Note that the local integrability conditions
($D$ -- simplex equations) we have not used at all, because the
diagonalization of our $U$ and the existence of a set of commutative
operators are not linked at all.
Nevertheless, the cases of the integrable models are more interesting
and we'll deal with the evolution operators $U$ constructed with
the help of $R$-s solving the Tetrahedron equation.

\section{Functional $R$ -- operator}

The interesting example of the Gaussian operators connected with
integrable models can be obtained from the functional $R$ --
operators
solving the functional tetrahedron equation (F.T.E.) \cite{kks-fte}.

Let a functional operator $R$:
\begin{equation}
\ds
R_{1,2,3}\,\circ\,[x_1,x_2,x_3]\;=\;[x'_1,x'_2,x'_3]\;,
\end{equation}
where
\begin{equation}
\ds
x'_1\;=\;r_1(x_1,x_2,x_3)\;,\;\;\;
x'_2\;=\;r_2(x_1,x_2,x_3)\;,\;\;\;
x'_3\;=\;r_3(x_1,x_2,x_3)\;,
\end{equation}
solves F.T.E. Examples of such $R$-s one can find in
Ref. \cite{oneparam}.
Mention now a very useful aspect of the functional maps. Namely, let
$x_k\in S_k$. Then actually $R$ can be
upgraded to $\hat R$ giving an isomorphism on a bundle of
$S_1\otimes S_2\otimes S_3$. Partially, on the tangent bundle,
\begin{equation}\label{diff}
\ds
\widehat R\,\circ\,[x_1,x_2,x_3|dx_1,dx_2,dx_3]\;=\;
[x'_1,x'_2,x'_3|dx'_1,dx'_2,dx'_3]\;,
\end{equation}
where
\begin{equation}\ds
dx'_i\;=\;\sum\;{\partial x'_i\over\partial x_k}\,dx_k\;,
\end{equation}
so as we can introduce the Gaussian operator
for the differentials on $S$:
\begin{equation}\label{gfun}
\ds
r\;=\;\widehat R\* R^{-1}\;,
\end{equation}
where, using the symbol $a_i$ instead of $dx_i$,
\begin{equation}
\ds
r\,\circ\,a_i\;=\;a'_i\;=\;\sum\;r_{i,k}(\vec x)\;a_k\;,\;\;\;
r_{i,k}(\vec x)\;=\;R^{-1}\,\circ\,{\partial x'_i\over\partial
x_k}\;.
\end{equation}
Thus for each functional operator
$R\;:\;S^{\otimes 3}\mapsto S^{\otimes 3}$ there appears
the gaussian action $G(r)\;:\;\sum dS\mapsto \sum dS$
for a still commutative algebra $dS$. $dS$ can be raised to
nontrivial $dS\oplus d^*S$,
\begin{equation}
\ds
r\,\circ\; a^*_i\;=\;\sum\; a^*_k\; (r^{-1})_{k,i}\;,
\end{equation}
with the conserving Weyl algebra $[a_i,a^*_k]\;=\;\delta_{i,k}$.

Because of $R$ solves F.T.E., $G(r)$ solves a local T.E. in the
direct
sum of $dS\oplus dS\oplus ...$. Thus the previous considerations give
the solution of the eigenvalue and eigenfunction problem for the
$2+1$ evolution governed by the properly quantized $R$.

Among other models one could mention several free bosonic or
fermionic
models, namely Bazhanov - Stroganov's free fermionic model and
Bazhanov - Baxter's free bosonic model.

\subsection{Bazhanov-Baxter's free bosonic model}

Recall in few words the form of $R$ -- operator connected with the
complex of the descendants of the Zamolodchikov -- Bazhanov -- Baxter
model
\cite{bb-first,bb-second,mss-vertex,fk-qd,br-qd,ms-modified}.
Let
\begin{equation}
\ds
u\*v\;=\;q\;v\*u\;,\;\;\;
w\;=\;-q^{-1/2}\,u\*v\;.
\end{equation}
Then for the quantum dilogarithm function \cite{fk-qd,br-qd}
\begin{equation}
\ds
\psi(x)\;=\;\prod_{n=0}^{\infty}\;(1-q^{1/2+n}\;x)
\end{equation}
over $u,v,w$ there exists the Pentagon equation
\begin{equation}\label{pentpsi}
\ds
\psi(v)\*\psi(u)\;=\;\psi(u)\*\psi(w)\*\psi(v)\;,
\end{equation}
and as the sequence of it the Tetrahedron equation for
\begin{equation}\label{rop}
\ds
R_{x,y,z}\;=\;
\psi(w_y^{-1}w_z^{})\,\psi(w_x^{-1}u_z^{})\,
\Pi_{x,y,z}\,
\psi(w_x^{}u_z^{-1})^{-1}\,\psi(w_y^{}w_z^{-1})^{-1}\;,
\end{equation}
\cite{ms-modified}, where
\begin{eqnarray}
&\ds
\Pi_{x,y,z}\,\circ\,\{v_x,v_y,v_z\}\;=\;\{v_x,v_xv_z,v_x^{-1}v_y\}\;,
&\nonumber\\
&\ds
\Pi_{x,y,z}\,\circ\,\{u_x,u_y,u_z\}\;=\;
\{u_xw_yw_z^{-1},v_x^{-1}u_z,v_xu_y\}\;,
&\nonumber\\
&\ds
\Pi_{x,y,z}\,\circ\,\{w_x,w_y,w_z\}\;=\;
\{w_xw_yw_z^{-1},w_z,w_y\}\;.
&
\end{eqnarray}
Profs. Bazhanov and Baxter in 1992 established the correspondence
between
the pair $u,v$ and the Gaussian exponents of the canonical Weil pair
$a,a^+$ \cite{bb-second}
\footnote{
This correspondence on the language of  $w$ -- functions on the
Fermat curves is given by
\begin{eqnarray}
&
\ds\sum_a\;\rightarrow\;\int\;da\;,\;\;\;\;
\ds\omega^{ab}\;\rightarrow\;\exp(-iab)\;,&\nonumber\\
&
\ds w(v,a)=\prod_{j=1}^a\;{\Delta(v)\over 1-v\omega^j}\;\rightarrow\;
\ds w(v,a)=\exp( -{i\over 2} {v\over 1-v} a^2)\;,&\nonumber\\
&
\ds\Phi(a)=(-)^a\omega^{a^2/2}\;\rightarrow\;
\ds\Phi(a)=\exp( -{i\over 2}a^2)\;,&
\nonumber\\
&
\ds\Delta(v)^N=1-v^N\;\rightarrow\;\Delta(v)=1-v\;.&
\end{eqnarray}
With this correspondence all the key relations in the Zamolodchikov
--
Bazhanov -- Baxter model conserve, $W$ -- type
model is well defined and its partition function is proportional
to the partition function of usual finite -- states ZBB model.}.
First, show as the Gaussian action like (\ref{gfun}) arises
in the language of the quantum dilogarithms. First, easily,
\begin{eqnarray}
\ds
\psi(u)\,\circ\, v & = & v\* (1\;-\;q^{1/2}\,u)^{-1}
\nonumber\\
\ds
\psi(v)\,\circ\, u & = & u\*(1\;-\;q^{-1/2}\,v)
\end{eqnarray}
In the limit when $q^{1/2}\rightarrow 1$ $\psi(u)$ and $\psi(v)$
become functional operators $\psi_f(u)$ and $\psi_f(v)$ and
\begin{equation}\label{luv}
\ds
\widehat\psi(u)_f\;=\;L_u\*\psi(u)_f\;,\;\;\;\;
\widehat\psi(v)_f\;=\;L_v\*\psi(v)_f\;,
\end{equation}
where for $\ds a^+\;=\;{du\over u}$ and
$\ds a\;=\;{dv\over v}$,
\begin{equation}
a\*a^+\;=\;a^+\* a \;+\; \lambda\;.
\end{equation}
'$+$' is not h.c., $q\;=\;\exp(\l)$, the Gaussian operators
\begin{equation}
\ds
G\;:\;\vectr{a}{a^+}\;\;\mapsto\;\;\vectr{a^\prime}{a^{+\prime}}\;=\;
\matr{\a}{\b}{\c}{\d}\*\vectr{a}{a^+}\;,
\end{equation}
where from the automorphism condition $[a',a^{\prime+}]\;=\l$
it follows that $\a\d-\b\c=1$, arise.
Eqs. (\ref{luv}) with
\begin{equation}
\ds
L_v\;=\;\exp({p\over 2\l}\; a^2)\;\;\;\mbox{ and }\;\;\;
L_u\;=\;\exp({q\over 2\l}\; a^{+2})
\end{equation}
give
\begin{equation}\label{up}
\ds
\exp ({p\over 2\l} a^2)\,\circ\,\vectr{a}{{a^+}}\;=\;
\matr{1}{0}{p}{1}\*\vectr{a}{{a^+}}\;,
\end{equation}
\begin{equation}\label{down}
\ds
\exp ({q\over 2\l} {a^+}^2)\,\circ\,\vectr{a}{{a^+}}\;=\;
\matr{1}{-q}{0}{1}\*\vectr{a}{{a^+}}\;.
\end{equation}
These operators should be completed by the diagonal one,
\begin{equation}\label{diag}
\ds
:\,\exp( {r\over \l} a^+\* a)\,:\,\circ\,\vectr{a}{a^+}\;=\;
\matr{(1+r)^{-1}}{0}{0}{(1+r)}\*\vectr{a}{a^+}\;,
\end{equation}
where the sign $::$ of the normal ordering is used: formally
\begin{equation}
\ds
:\,\exp( {r\over \l} a^+\*a)\,:\;=\;
\sum_{n=0}^\infty\;
{(r/\l)^n\over n!}\;a^{+n}\*a^n\;.
\end{equation}
These three operators give the Gaussian triangular expansion of
$CP(2)$. Mention one more operator,
\begin{equation}\label{extra}
\ds
\exp ({r\over 2\l} (a+a^+)^2)\,\circ\,\vectr{a}{{a^+}}\;=\;
\matr{1-r}{-r}{r}{1+r}\*\vectr{a}{{a^+}}\;.
\end{equation}
The exponential forms (these Gaussian exponents
define our terminology) are rather formal, because their explicit
forms
depend actually on a representation, and the following integral form
is
more useful sometimes then the series decomposition:
\begin{equation}\label{repr}
\ds\exp({p\over 2\l}\, a^2)\;=\;
\int\;d\zeta\,\exp({\l p\over 2}\,\zeta^2\,+\,i\,p\,\zeta\,a)
\end{equation}
etc.

There are two remarkable relations for these Gaussian
operators: \\first, Kashaev's relation \cite{rmk-lybe}:
\begin{equation}
\exp({x\over 2}a^2)\*\exp({y\over 2}{a^+}^2)\*\exp({z\over
2}a^2)\;=\;
\exp({z'\over 2}a^{+2})\*\exp({y'\over 2}a^2)\*\exp({x'\over
2}a^{+2})\;,
\end{equation}
where the electric network transformation arises
\begin{eqnarray}
\ds x' & = &{x\,y\over x\,+\,z\,-\,\l^2\, x\,y\,z}\;,
\nonumber\\
\ds y' & = &x\,+\,z\,-\,\l^2\, x\,y\,z\;,
\nonumber\\
\ds z' & = &{y\,z\over x\,+\,z\,-\,\l^2\,x\,y\,z}\;.
\end{eqnarray}
Second, Bazhanov's ``restricted star -- triangle''
for the bosonic $3D$ model \cite{bb-second}:
\begin{equation}\label{pentagon}
\exp({ p\over 2\l}a^2)\*
\exp({ q\over 2\l}a^{+2})\;=\;
\exp({q'\over 2\l}a^{+2})\*
\exp({r'\over 2\l}(a^++a)^2)\*
\exp({p'\over 2\l}a^2)\;,
\end{equation}
where
\begin{eqnarray}
\ds
q' & = & q\;{1-p\over 1-pq}\;,
\nonumber\\
\ds
p' & = & p\;{1-q\over 1-pq}\;,
\nonumber\\
\ds
r' & = & pq\;.
\end{eqnarray}
This is the usual pentagon relation connected with eq.
(\ref{pentpsi}).
Although we can't extract $\exp a^2$ directly from
the quantum dilogarithm, the identification of the spectral
parameters of exp-s with the functional transformation
generated by the quantum dilogarithm $\psi(v)$ when
$q^{1/2}\rightarrow 1$ and $v\mapsto v\,\exp(a)$, following from
eq. (\ref{luv}), gives the correspondence
\begin{equation}\label{correspondence}
\ds
\psi(v)\rightarrow \;[\mbox{ Func. Part }\psi_f(u)]\,\*\,
\exp(-{1\over 2\l}{v\over 1-v}\, a^2)\;,
\end{equation}

Give just an example a representation of eq. (\ref{pentagon}). Let
$\l$ be pure imaginary so that the pair $a$ and $a^+$ is
the pair {\em coordinate -- momentum}. Take the representation
when
\begin{eqnarray}
&\ds
a\* |x>\;=\;
(\l\,x\,-\,{\partial\over\partial x})\,|x>\;,\;\;\;\;
a^+\* |x>\;=\;
{\partial\over\partial x}\,|x>\;,&\nonumber\\
&\ds
<x|\* a\;=\;
<x|\,\stackrel{\leftarrow}{{\partial\over\partial x}}\;,\;\;\;\;
<x|\* a^+\;=\;
<x|\,(\l\,x\,-\,\stackrel{\leftarrow}{{\partial\over\partial x}})\;,&
\end{eqnarray}
so that
\begin{equation}
\ds
<y|x>\;=\;\exp({\l\over 2}\,x^2)\;\d (x-y)\;,
\end{equation}
and
\begin{equation}
\ds 1\;=\;\int\; |x>\,\exp(-{\l\over 2}\,x^2)\,dx\,<x|\;.
\end{equation}
Then
\begin{eqnarray}
\ds
<x|\exp({p\over 2\l}\,a^2)|y> & = &
\exp(-{\l\over 2p}\,(x-y)^2\,+\,
{\l\over 2}\,y^2)\;,\nonumber\\
\ds
<x|\exp({q\over 2\l}\,a^{+2})|y> & = &
\exp(-{\l\over 2q}\,(x-y)^2\,+\,
{\l\over 2}\,x^2)\;,\nonumber\\
\ds
<x|\exp({r\over 2\l}\,(a^++a)^2)|y> & = &
\exp({\l\over 2}\,(1+r)\,y^2)\,\delta (x-y)\;.
\end{eqnarray}
In this representation there appears exactly Bazhanov -- Baxter's
bosonic restricted star -- triangle relation.

Return now to $R$ -- matrix (\ref{rop}). We'll identify
\begin{equation}
\ds
v\rightarrow v\,\exp(a-a^+)\;,\;\;\;
u\rightarrow u\,\exp(a^+)\;,\;\;\;
w\rightarrow w\,\exp(a)\;.
\end{equation}
We'll deal with the endomorphisms of the direct sum of several
pairs of
\begin{equation}\label{weil}
\ds
\vec a_k \;=\; \vectr{a_k}{a^+_k}\;,
\end{equation}
The equivalence between $\psi$ and $\exp a^2$ is given by
(\ref{correspondence}).

As it was described in the previous sections, for the $R$ -- matrix
being treated as the Gaussian operator $R=G(r)$, we are interested in
the
matrix $r$ and the evolution governed by this $r$. Thus for
$\Pi_{x,y,z}$
\begin{equation}
\ds
\pi_{x,x}\;=\;\pi_{y,z}\;=\;\pi_{z,y}\;=\;1\;,
\;\;\;\;
\pi_{y,y}\;=\;\pi_{z,z}\;=\;0\;,
\end{equation}
and
\begin{equation}
\ds
\pi_{x,y}\;=\;-\pi_{x,z}\;=\;\matr{1}{0}{1}{0}\;,\;\;\;\;
\pi_{y,x}\;=\;-\pi_{z,x}\;=\;\matr{0}{0}{-1}{1}\;.
\end{equation}
Restore all other exp-s:
\begin{eqnarray}
&\ds
R_{x,y,z}\;=\;
\exp({p_1\over 2\l}(a_y-a_z)^2)\*
\exp({p_2\over 2\l}(a_x-a_z^+)^2)\*
&\nonumber\\
&\ds
\pi_{x,y,z}\*
\exp(-{p_3\over 2\l}(a_x-a_z^+)^2)\*
\exp(-{p_4\over 2\l}(a_y-a_z)^2)\;,&
\end{eqnarray}
where from eq. (\ref{rop}) it follows that
\begin{equation}
\ds
p_1(1-p_2)(1-p_4)\;=\;p_4(1-p_3)(1-p_1)\;.\label{ccc}
\end{equation}
Two matrix functions arise:
\begin{equation}
\ds
\exp({p\over 2\l}\,(a_x-a_z^+)^2)
\;\rightarrow\; \sigma(p)
\end{equation}
and
\begin{equation}
\ds
\exp({p\over 2\l}\,(a_y-a_z)^2)
\;\rightarrow\; \sigma'(p)\;.
\end{equation}
Then
\begin{equation}
\ds
\sigma (p)\;=\;
\left(\begin{array}{rrrrrr}
1 & 0 &   0 & 0 &   0 &  0 \\
p & 1 &   0 & 0 &   0 & -p \\
   0 &   0 & 1 & 0 & 0 &  0 \\
   0 &   0 & 0 & 1 & 0 &  0 \\
p & 0 &   0 & 0 &   1 & -p \\
0 & 0 &   0 & 0 &   0 &  1 \end{array}\right)\;,
\end{equation}
and
\begin{equation}
\ds
\sigma' (p)\;=\;
\left(\begin{array}{rrrrrr}
 1 & 0 &   0 & 0 &   0 & 0 \\
 0 & 1 &   0 & 0 &   0 & 0 \\
 0 & 0 &   1 & 0 &   0 & 0 \\
 0 & 0 &   p & 1 &  -p & 0 \\
 0 & 0 &   0 & 0 &   1 & 0 \\
 0 & 0 &  -p & 0 &   p & 1 \end{array}\right)\;,
\end{equation}
and
\begin{equation}\label{rssss}
\ds
r\;=\; \sigma'(-p_4)\* \sigma(-p_3)\* \pi_{x,y,z}\*
\sigma(p_2)\* \sigma'(p_1)\;.
\end{equation}
Thus the eigenvector problem for the bosonic model reduces
to the six - dimensional eigenvector problem. Let amplitudes for
eq. (\ref{weil}) be
\begin{equation}
\ds
\vec x^T_p\;=\;(x_p\,,\;x^*_p)\;\;\;\mbox{etc.}
\end{equation}
(do not mix $p$ as a point on the lattice with $p$ as a parameter).
Then the amplitudes of the eigenvectors are given by eq.
(\ref{eigen}),
where the final expression for $r$ (\ref{rssss}) is
\begin{equation}
\ds
r\;=\;\left(\begin{array}{cccccc}
1-P_1 &     0 &   Q_2 &     0 &  -Q_2 &   P_1 \\
    0 & 1-Q_1 &   P_3 &   Q_1 &  -P_3 &     0 \\
  P_1 &     0 & 1-Q_2 &     0 &   Q_2 &  -P_1 \\
 -Q_3 &   P_2 &     0 & 1-P_2 &     0 &   Q_3 \\
    0 &  -Q_1 &   P_3 &   Q_1 & 1-P_3 &     0 \\
  Q_3 &  -P_2 &     0 &   P_2 &     0 & 1-Q_3
\end{array}\right)\;,\label{exmat}
\end{equation}
\begin{eqnarray}
&\ds
P_1\;=\; p_2\;,\;\;\; P_2\;=\;1-p_3p_4\;,\;\;\;
P_3\;=\;1-p_3(1-p_1)\;,&
\nonumber\\&\ds
Q_1\;=\; p_3\;,\;\;\; Q_2\;=\;1-p_1p_2\;,\;\;\;
Q_3\;=\;1-p_2(1-p_4)\;.
\end{eqnarray}
The characteristic polynom (\ref{chp})
for $r$ -- matrix (\ref{rssss}) does not factorize except some
special
limits, and the direct calculations for the integral (\ref{integral})
give the expression
\begin{equation}\label{pf}
\ds
z_0(p_1,p_2,p_3,p_4)\;=\; B(p_1)\; B(p_2)\; B(p_3)\; B(p_4)\;,
\end{equation}
where
\begin{equation}\label{B}
\ds
B(p)\;=\;\exp {2 i\over \pi} \varepsilon(p)\left(
\mbox{Li}(p)-\mbox{Li}(1/2)\right)\;,
\end{equation}
\begin{equation}
\ds
\varepsilon(p)\;=\;\mbox{sign Im} (p)\;,
\end{equation}
and Roger's dilogarithm
\begin{equation}
\ds
\mbox{Li}(p)\;=\;-\;{1\over 2}\;\int_0^p\;\left(
{\log x\over 1-x}\;+\;{\log 1-x\over x}\right)\;d\;x\;.
\end{equation}
These calculations are cumbersome, but they can be simplified
significantly in the limit when $p_1=p_4=0$. In this case the
characteristic polynom factorizes and there appeared the integral
definition of $B$:
\begin{eqnarray}
&\ds
B(p)\;=\;\exp -{1\over (2\pi)^3}\;
\int\int\int_0^{2\pi}\;
d\phi_1\,d\phi_2\,d\phi_3\*&\nonumber\\
&\ds
\log(
1-p(e^{i\phi_1}+e^{i(\phi_2+\phi_3)})
-(1-p)(e^{i\phi_2}+e^{i(\phi_1+\phi_3)})
+e^{i(\phi_1+\phi_2+\phi_3)})
&\nonumber\\
&&\label{def}
\end{eqnarray}
This integral can be taken easily and gives (\ref{B}).
Match the expression for $z_0$ with Baxter's results for
Zamolodchikov
model $z_2$ \cite{baxter-pf} and Bazhanov's result for the bosonic
model
$z_B$ \cite{bb-second}:
\begin{equation}
z_0\;=\;z_2^{-4}\;=\; z_B^2.
\end{equation}
The origin of the difference between $z_0$ and Bazhanov's $z_B$
is that up to the proper representation our system of the
eigenvectors
is overdefined twice (see the remark at the end of Section 3).

Few words concerning the eigenvectors of $U$. First, the
characteristic
polynom (\ref{chp}) for $r$ (\ref{exmat}) has the following property:
\begin{equation}
\ds
\chi(x,y,z|r)\;=\;x^2\;y^2\;z^2\;\chi(x^{-1},y^{-1},z^{-1}|r)\;.
\end{equation}
Thus for the isotropic eigenvectors $x=y=z$ six roots of
$\chi=0$ divide into two subsets:
\begin{equation}\ds
\{\lambda_1,\;\lambda_2,\;\lambda_3\}\;\;\;
\mbox{and}\;\;\;
\{\lambda_4=\lambda_1^{-1},\;
\lambda_5=\lambda_2^{-1},\;
\lambda_6=\lambda_3^{-1}\}\;,
\end{equation}
shuch that $|\lambda_{1,2,3}|<1$ (note, $|\lambda_k|=1$ only if
all $p$-s are real). This property remains for arbitrary
quasimomenta $k_\a,k_\b$:
\begin{equation}\ds\label{lll}
|\lambda_{1,2,3}(k_\a,k_\b)|<1\;\;\;
\mbox{and}\;\;\;
\lambda_{4,5,6}(k_\a,k_\b)\;=\;
\lambda_{1,2,3}(k_\a^{-1},k_\b^{-1})^{-1}\;.
\end{equation}
In order to obtain the convergency of $\mbox{Trace}\;U^M$, suppose
the eigenvectors for the linear evolution operator
$\omega_{1,2,3}(k_\a,k_\b)$ to be the creation operators. so the
vacuum vector is to be defined
\begin{equation}
\ds
\omega_{4,5,6}(k_\a,k_\b)\;|\Omega>\;=\;0\;.
\end{equation}
Due to Eq. (\ref{lll}) the structure of
the bulk free energy formula (\ref{integral}) for $k_0$
does not change, but the extra  multiplier ${1\over 2}$
appears: $\ds k\;=\;{1\over 2}\;k_0$.
Recall, in the case then $p_1,p_2,p_3,p_4$ are real, all $\lambda$-s
belong to the unit circle, hence when one changes the signs of
$\mbox{Im}\,p$, the structure of the physical vacuum changes,
so $B(p)$ depends on the sign of $\mbox{Im}\,p$ so drastically.

Note, the derivation of the partition function via the three
dimensional
integral was suggested originally for the free bosonic model
by V. V. Bazhanov \cite{vvb-pc}.

\section{Summary}

In this paper we have formulated the evolution problem for
$D+1$ dimensional free models. The evolution $U$ -- type operators
act linearly on the algebra of the creation and annihilation
operators,
and so  due to the locality of $U$ the eigenvector problem can be
solved
directly up to a finite eigenvector problem. Appeared physical
creation
and annihilation operators depending on $D$ momenta diagonalize
the evolution operator. In the case of integrable models,
when there exists a set of $T$ -- type
transfer matrices commuting with $U$, the eigenvectors
diagonalize these $T$ as well. The existence of the set $T$ for
the given $U$ is provided by the $D+1$ simplex equation
\footnote{Note that one can construct a $T$ -- type commutative set
without $D+1$ simplex equation for $D>1$. This fact was mentioned
by Bazhanov and Stroganov in Ref. \cite{bs-3dffm}}.

\vspace{1cm}

\noindent
{\large\bf Acknowledgments}:
We should like to thank Yu. G. Stroganov, H. E. Boos,
V. V. Mangazeev, G. P. Pron'ko and  M. V. Saveliev
for many fruitful discussions.


\end{document}